\documentclass[sn-aps]{sn-jnl}% American Physical Society (APS) Reference Style

\jyear{2022}%

\theoremstyle{thmstyleone}%
\theoremstyle{thmstyletwo}%

\theoremstyle{thmstylethree}%

\begin{document}

\title[Tensor-to-Scalar Transition Predictions in CLAS12]{Predicted Measurements of the Tensor-to-Scalar Transition in the CLAS12 Nuclear Targets Experiment} 
\author*[1]{\fnm{Erin Marshall} \sur{Seroka}}\email{erinseroka@gwmail.gwu.edu}

\author[1]{\fnm{Axel} \sur{Schmidt}}\email{axelschmidt@gwu.edu}

\affil*[1]{\orgdiv{Physics Department}, \orgname{The George Washington University}, \orgaddress{\street{725 21st Street NW}, \city{Washington},  \state{DC} \postcode{20052}, \country{USA}}}

\abstract{Short-range correlated (SRC) nucleon pairs, which are strongly interacting nucleons at short inter-particle distances, can reveal properties of the effective nucleon-nucleon (\textit{NN}) interaction at short distance scales. The relative abundance of proton-proton (\textit{pp}) pairs and proton-neutron (\textit{pn}) pairs, for example, is sensitive to the tensor contribution to the \textit{NN} interaction. Generalized Contact Formalism (GCF) theory, when used with realistic phenomenological \textit{NN} potential models, predicts a transition from a tensor-dominated regime---at relative momenta of approximately 400~MeV/\textit{c} where \textit{pp} pairs are suppressed relative to \textit{pn} pairs---to a scalar-dominated regime at higher momenta with no preferred isospin projection. While an increase in the prevalence of \textit{pp} pairs with increasing momentum has been observed in a few experiments, difficulties associated with neutron detection have so far hindered the observation of a corresponding reduction in the abundance of \textit{pn} pairs. High-precision measurements showing a simultaneous increase in the abundance of \textit{pp} pairs and change in the abundance of \textit{pn} pairs with increasing momentum would conclusively demonstrate the existence of the tensor-to-scalar transition. In this work, we study the potential impact of the recently conducted Nuclear Targets Experiment at the CLAS12 detector at Jefferson Lab, using GCF simulations. We model the expected yields and relevant observables for a carbon target with a beam energy of 6 GeV and show that sufficient statistical precision can be obtained from the experimental data, both for \textit{pp} and \textit{pn} pairs, to observe the tensor-to-scalar transition.}

\keywords{Short-Range Correlations, NN Interaction, CLAS12, Generalized Contact Formalism, Quasielastic Electron Scattering}

\maketitle

\section{Introduction}\label{sec1}

Experiments and \textit{ab initio} calculations in recent years have shown that, in nuclei heavier than carbon, about 20\% of nucleons occupy high-momentum states above the Fermi momentum $k_{F}$~\cite{Fomin:2011ng,Schmookler2019,Wiringa:2013ala,Lonardoni:2017egu,Lonardoni:2017hgs}, which consist almost entirely of pairs of strongly interacting nucleons in close proximity.  
These short-range correlated (SRC) pairs move through the nucleus with a small center-of-mass momentum, typically below $k_F$, that is comparable to the momenta of slower, uncorrelated nucleons~\cite{Cohen2018}.
SRC pairs may play a significant role in outstanding questions in nuclear and particle physics, including the nuclear matrix elements in double beta decay~\cite{Kortelainen2007,Simkovic2009,Benhar2014}, nuclear charge radii \cite{Miller2019}, and the EMC effect~\cite{Schmookler2019}. Recent reviews include Refs.~\cite{Arrington:2022sov,Hen:2016kwk}.

One of the striking features of short range correlations is the strong prevalence of neutron-proton ($np$) pairs, i.e., an isospin-0, spin-1, configuration, over proton-proton ($pp$) or neutron-neutron ($nn$) pairs, i.e., any isospin-1 configuration. 
This property, referred to as $np$-dominance, has been observed through a wide-range of techniques, including proton scattering reaction~\cite{Piasetzky:2006ai}, exclusive electron scattering~\cite{JeffersonLabHallA:2007lly,Subedi:2008zz,LabHallA:2014wqo,Hen:2014nza,DuerPRL,CLAS:2020mom,Korover2021}, and most recently through inclusive electron scattering on isospin mirror nuclei~\cite{Li:2022fhh}. 
Studies of asymmetric nuclei have shown that $np$-dominance persists even when there is an imbalance between proton and neutron numbers~\cite{Hen:2014nza,DuerPRL,Li:2022fhh}, which has implications for understanding the structure of neutron-rich matter like that found in the cores of neutron stars \cite{Li2018,Frankfurt2008}. 

In a nonrelativistic picture of nucleon-nucleon ($NN$) pairs, the origin of $np$-dominance lies in the tensor part of the $NN$ interaction, which favors a spin-1 configuration. At distance scales where the central part of the $NN$-interaction transitions from long-range attraction to short-range repulsion---on the order of the separation distance of nucleons in SRC pairs---the tensor component plays a significantly larger role than the scalar component. Theory calculations with realistic $NN$ potentials show that SRCs are dominated by the spin-1 isospin singlet of $pn$ pairs and the spin-0 isospin triplet of $pp$, $pn$, and $nn$ pairs. For two nucleons in a spin-0, isospin-1 configuration, including $pp$ pairs, the two-body momentum distribution has a node in the momentum range of 300--600~MeV/$c$, while the spin-1, isospin-0 configuration does not \cite{CiofidegliAtti2015,Ryckebusch2019,Sargsian2005,Schiavilla2007,Alvioli2008,Weiss2018}. Therefore, the abundance of $pp$ pairs is most suppressed relative to $pn$ pairs at approximately 400~MeV/$c$ due to the preference for spin-1 $pn$ pairs by the tensor part of the $NN$ interaction~\cite{Weiss2018}. A dramatic increase can be expected in the abundance of $pp$ pairs from momentum of about 400~MeV/$c$ to momentum of about 1000~MeV/$c$. Simultaneously, the abundance of $pn$ pairs decreases in the tensor-to-scalar transition from approximately 400~MeV/$c$ to about 1000~MeV/$c$. In the high-momentum limit, where the scalar part of the $NN$ interaction dominates over the tensor part, the abundance of pairs is expected to be equal for all Pauli-allowed spin-isospin projections.

The tensor-to-scalar transition can be investigated through scattering experiments in which a probe interacts with a correlated nucleon in a target nucleus, inducing a hard break-up of the pair. Though several different probes have been used, in this work we focus on electron scattering. We specifically consider quasielastic electron scattering reactions with large momentum transfer, in which the struck nucleon is a proton from an SRC pair, with cross section $(e,e'p)$. We refer to the outgoing struck proton as the ``leading proton,'' since at high momentum transfer it leaves the nucleus with significantly higher momentum than the recoil nucleon from the same correlated pair. Contributions from reactions with non-SRC leading protons, such as multi-nucleon and meson-exchange currents and final state interactions (FSI), can be suppressed by a set of kinematic criteria, including requiring that the initial proton momentum be above the Fermi momentum $k_F$ and requiring high momentum transfer~\cite{Hen:2014nza,DuerPRL,CLAS:2020mom,Korover2021}. As a subset of the $(e,e'p)$ reaction channel, we look for exclusive events with a recoiling spectator nucleon that originated as the correlated partner to the struck proton. These events have cross section $(e,e'pN)$, where the recoil nucleon $N$ may be either a proton ($p$) or a neutron ($n$).

As shorthand, we define $pp/p$ and $pn/p$ to be the cross section ratios $(e,e'pp)/(e,e'p)$ and $(e,e'pn)/(e,e'p)$, respectively. The tensor-to-scalar transition can be mapped by observing changes in $pp/p$ and $pn/p$ as a function of missing momentum. The missing momentum is defined to be $\vec{p}_{\rm{miss}}=\vec{p}_{1}-\vec{q}$, where $\vec{p}_{1}$ is the measured outgoing momentum of the leading proton and $\vec{q}$ is the momentum transferred by the electron. Under the plane wave impulse approximation, $\vec{p}_{\rm{miss}}$ is a proxy for the initial momentum, $\vec{p}_i$, of the struck proton.
The tensor-to-scalar transition has been successfully demonstrated through the $pp/p$ channel \cite{CLAS:2020mom}. 
However, measurements of $pp$ are susceptible to inflation due to single charge exchange (SCX) reactions of $pn$ or $np$ pairs. 
Events categorized as $pp$ may have originated as $pn$ pairs in which the neutron swapped isospin projections with a proton through the exchange of a charged pion. 
This is especially concerning since $pn$ pairs are an order of magnitude more abundant than $pp$ pairs at momenta near 400 MeV/$c$. Therefore, observing an increase in $pp/p$ alone is not sufficient to conclusively demonstrate the transition from a tensor-dominated interaction to a scalar-dominated interaction.

In the cross section ratio $pn/p$, SCX contributions from proton-to-neutron and neutron-to-proton interactions are expected to roughly cancel. 
$pn/p$ is a complementary measurement to $pp/p$, which, although more ambitious due to the experimental challenges of neutron detection, is crucial because it is significantly less sensitive to contamination from SCX reactions.

Two previous studies extracted $pn/p$ from quasielastic electron scattering data in helium-4 \cite{LabHallA:2014wqo} and carbon-12~\cite{Korover2021}, but unfortunately, their large uncertainties precluded any definite conclusions about the onset of the repulsive core from this channel. In fact, in Ref.~\cite{Korover2021}, the measurement is consistent with a constant value of $pn/p$. To date, no analysis has reported simultaneous measurements of $pp/p$ and $pn/p$ with a statistically significant change in both quantities across the momentum range. A higher statistics measurement, made possible by the increased rate capabilities of the CLAS12 Spectrometer and a new dedicated neutron detection system~\cite{Chatagnon:2020lwt}, could provide confirmation of the tensor-to-scalar transition, test the role played by single charge exchange reactions, and test the effectiveness of different $NN$ interaction models with the Generalized Contact Formalism.

\section{The CLAS12 Nuclear Targets Experiment}
\label{sec:clas12}

\begin{table}[thb!]
\centering
\caption{\label{tab:rgm-data} CLAS12 Nuclear Targets Experiment Data}
\begin{tabular}{|c|c|c|c|}
\hline
Target & Energy (GeV) & Integrated Luminosity (fb$^{-1}$) & Total Charge (mC) \\
\hline
H & 5.99 & 3.68 & 2.92\\
H & 2.07 & 0.39 & 0.31\\
D & 5.99 & 52.00 & 16.87\\
D & 2.07 & 1.81 & 0.59\\
He & 5.99 & 57.24 & 24.37\\
C & 5.99 & 80.31 & 48.56\\
C & 4.03 & 10.46 & 6.33\\
C & 2.07 & 1.00 & 0.60\\
Ar & 5.99 & 4.91 & 1.87\\
Ar & 4.03 & 6.67 & 2.54\\
Ar & 2.07 & 1.35 & 0.51\\
$^{40}$Ca & 5.99 & 51.59 & 44.28\\ 
$^{48}$Ca & 5.99 & 20.75 & 17.81\\
Sn & 5.99 & 9.29 & 12.06\\ 
\hline
\end{tabular}
\end{table} 

Jefferson Lab Experiment E12-17-006A, titled, ``Exclusive Studies of Short Range Correlations in Nuclei using CLAS12,'' and hereafter referred to as the CLAS12 Nuclear Targets Experiment, collected data between November 2021 and February 2022 at the Thomas Jefferson National Accelerator Facility (or ``Jefferson Lab") in Newport News, Virginia.

\begin{figure}[h]%
\centering
\includegraphics[width=0.7\textwidth]{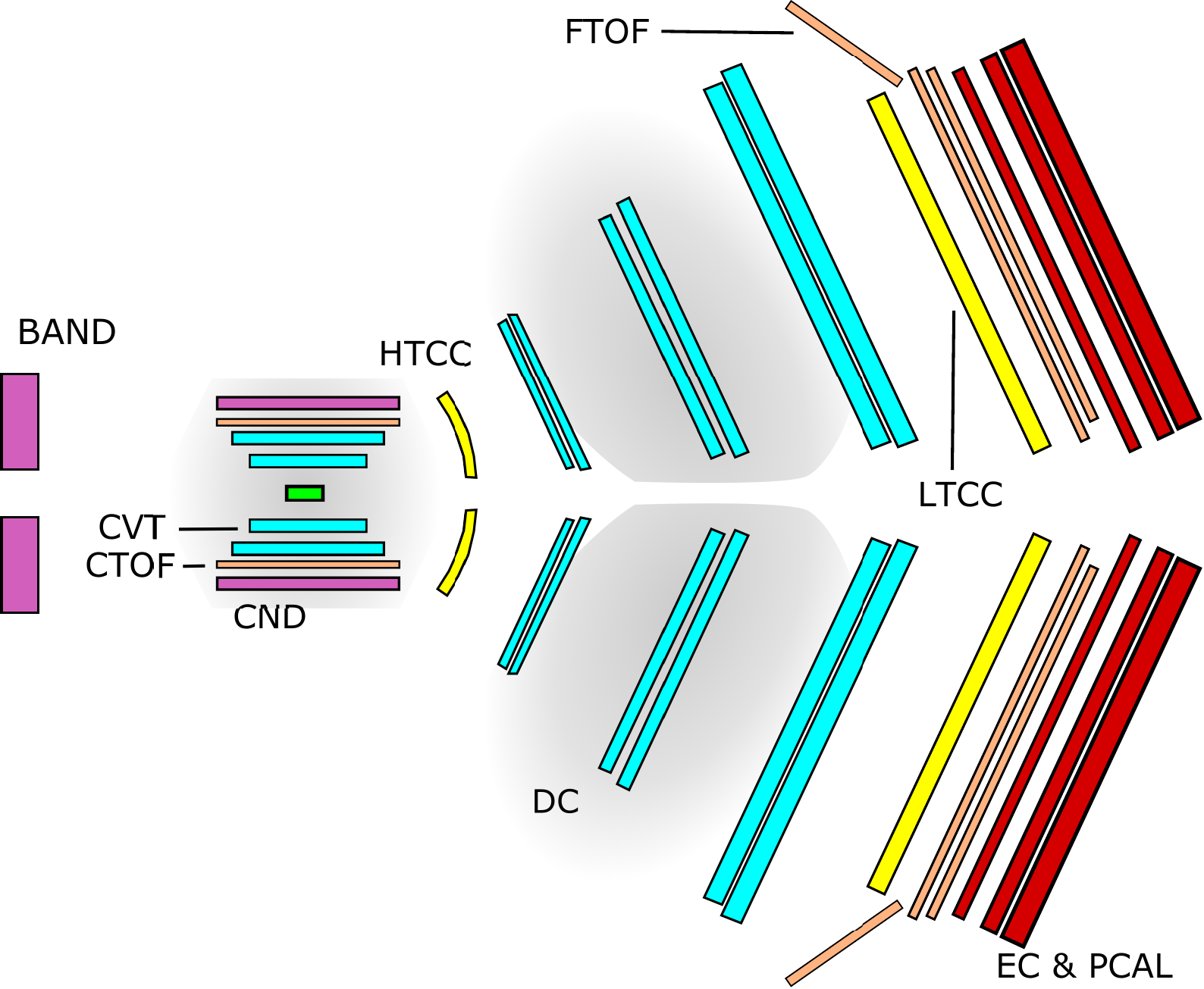}
\caption{A schematics cross section of the CLAS12 detector, approximately to scale. The Central Detector wraps around the target (green) and includes the Central Vertex Tracker (CVT, blue), Central Time-of-Flight Detector (CTOF, orange), and Central Neutron Detector (CND, magenta). The Forward Detector, arranged in six sectors around the beam line, includes the High Threshold Cherenkov Counter (HTCC, yellow), Drift Chambers (DC, blue), Low Threshold Cherenkov Counter (LTCC, yellow), Forward Time-of-Flight Detector (FTOF, orange), and the Electromagnetic Calorimeter and Preshower Calorimeter (EC \& PCAL, red). Also shown is the Backward Angle Neutron Detector (BAND, magenta). The gray regions indicate the magnetic fields of the solenoid (left) and torus (right) magnets.}
\label{clas12}
\end{figure}

The experiment was conducted as part of the CLAS Collaboration's Run Group M, during which an electron beam at an energy of 2, 4, and 6 GeV, was scattered from $^{1}$H, $^{2}$H, $^4$He, C, $^{40}$Ca, $^{48}$Ca, Ar, and Sn targets. For the 6 GeV beam energy, a four foil carbon target with a total thickness of 2 mm was used. Outgoing particles were detected by the CLAS12 Spectrometer~\cite{nima-clas12}.
This experimental run was the first use of nuclear targets heavier than deuterium with CLAS12. Information about the data accumulated on all targets is given in Table \ref{tab:rgm-data}. The primary goal of the CLAS12 Nuclear Targets Experiment was to perform high-precision and exploratory SRC studies on $^4$He and C and to compare SRC characteristics across a range of nuclei. 
More information can be found in the experiment's proposal~\cite{rgm-proposal}, although the plan and scope of the experiment was updated somewhat between the proposal and data collection.

CLAS12 is a large-acceptance magnetic spectrometer situated in Jefferson Lab's Experimental Hall B. 
CLAS12 is the successor to the retired CLAS detector~\cite{CLAS:2003umf}, which provided the data for several ground-breaking findings in the study of short-range correlations \cite{DuerPRL,Duer2018,CLAS:2020mom,Hen:2014nza,Cohen2018,Schmookler2019,Korover2021}. 
CLAS12 is arranged with a Forward Detector with a toroidal magnetic field and a Central Detector with a barrel-shaped solenoid magnet surrounding the target, as shown in Fig.~\ref{clas12}.
The Forward Detector, which includes some sub-detectors from the previous CLAS Spectrometer, covers scattering angles of approximately
$5^\circ < \theta < 40^\circ$, and includes drifts chambers for charged-particle tracking~\cite{Mestayer:2020saf}, Cherenkov detectors for particle identification~\cite{Sharabian:2020whm,Contalbrigo:2020lnd}, time-of-flight scintillation counters~\cite{Carman:2020fsv}, and an electromagnetic calorimeter~\cite{Asryan:2020iqj}. The Central Detector, an entirely new compact barrel-shaped set of detectors, extends the polar angle acceptance to approximately $40^{\circ} < \theta < 140^{\circ}$. The Central Detector includes components for charged particle tracking and timing~\cite{Acker:2020qkv,Carman:2020yma}, as well as a dedicated Central Neutron Detector (CND)~\cite{Chatagnon:2020lwt}.

The CND allows detection of neutrons with momenta from approximately 0.2 to 1.0 GeV/$c$ with a maximum efficiency of about 10\%~\cite{nima-clas12}. Precise measurements of the momentum and polar angle dependence of the neutron detection efficiency for the Nuclear Targets Experiment are underway. Neutron detection in the former CLAS detector was accomplished using either the electromagnetic calorimeter, with efficiency in the range of 35-60\%~\cite{Amarian_EC_nim,lachinet.gmn,lachinet.thesis,Duer2018}, or the time-of-flight detector, with efficiency of 2-8\%~\cite{lachinet.thesis,Korover2021}. Simulations have indicated that under the conditions of the Nuclear Targets Experiment, more recoil nucleons can be expected in the Central Detector than in the Forward Detector, and very few recoil nucleons can be expected in the Backward Angle Neutron Detector (BAND), making the CND the optimal choice for neutron detection in this study.

The enhanced capabilities of the CLAS12 Spectrometer have the potential to lead to significant progress on a number of open questions related to short-range correlations. First, CLAS12 can accommodate a luminosity of $10^{35}$~cm$^{-2}$s$^{-1}$, a factor of 10 higher than the original CLAS, which is valuable for improving the precision on relatively rare quasielastic knockout of correlated nucleons, such as those at high $p_{\rm{miss}}$. Second, CLAS12 has a larger angular acceptance than CLAS, which is valuable for the reconstruction of exclusive multi-nucleon final states. Third, CLAS12's Central Neutron Detector provides neutron detection capabilites over a wide solid angle (approximately polar angles of $40^{\circ}$ to 140$^{\circ}$), which is especially valuable for the study described in this work. The combination of these advantages will drastically improve searches for 3-nucleon short-range correlations, for example~\cite{rgm-proposal}.

For our study, we specifically consider the portion of the experiment with a 6~GeV beam and a carbon target as this setting has the largest number of events for the quasielastic SRC break-up reactions of interest, and for comparison to Ref.~\cite{Korover2021}. The same analysis could be extended to other beam/target combinations.

\section{Simulation}

\label{sec:sim}

To investigate the expected measurements of the tensor-to-scalar transition in the CLAS12 Nuclear Targets Experiment, we have performed simulation studies of the experimental conditions using the theoretical predictions of Generalized Contact Formalism (GCF), a factorized model of short-range correlations, according to Ref.~\cite{Pybus2020}. Our simulated events contain a scattered electron from a 6 GeV beam, a leading proton (assumed to have interacted with the electron through a virtual photon), and a recoil nucleon (assumed to be the correlated spectator nucleon in an SRC pair) from a $^{12}$C nucleus. We have studied the events both by integrating over the approximate experimental acceptance (so-called ``Fast Monte Carlo''), as well as by using the Geant4 simulation package and CLAS12 reconstruction programs to produce simulated data to analyze in the same manner as the experimental data. The Geant4 simulation includes a detailed model of the experimental response, including detector efficiency and resolution. The Fast Monte Carlo, on the other hand, assumes perfect efficiency within the acceptance as well as perfect resolution. By examining both, we can study the impact of detector resolution and efficiency on the observables of interest.

We generated Monte Carlo events according to the theoretical cross sections of Generalized Contact Formalism, a factorized, scale-separated model of short-range correlations among nucleons. Theoretical studies have long demonstrated that high-momentum nucleons (with momenta exceeding the Fermi momentum $k_F$) are dominated by short-range correlated nucleon pairs, whose behavior can be described using a factorized wave function based on the relative and center-of-mass momentum distributions of the pairs~\cite{Frankfurt1981,CiofiDegliAtti1991}. Some of the hallmark features of SRC behavior, including the scaling for ratios of inclusive cross sections of nuclei to the deuteron at $x_{B}>1$~\cite{Frankfurt1993}, where $x_B$ is the Bjorken scaling parameter defined in Section~\ref{ssec:fastmc}, were predicted in these early works and later demonstrated experimentally at facilities like SLAC and Jefferson Lab. The development of Generalized Contact Formalism drew upon the aforementioned work and upon a contact formalism used to analyze the properties of ultracold Fermi gases~\cite{tan1,tan2,tan3}. 
Detailed information about the development of GCF and its application to nucleon-nucleon correlations can be found in Refs.~\cite{Weiss2015,Weiss2018,Weiss2019,Pybus2020} and is briefly summarized here. 

GCF predicts that, at high missing momentum, the $(e,e'pN)$ scattering cross section approximately factorizes according to three distinct momentum scales:
the high-momentum scale, defined by the momentum transferred to the nucleus by the probe, typically greater than 1~GeV/$c$; the medium-momentum scale, describing the relative momentum between the correlated nucleons, approximately $300$--$1000$~MeV/$c$; and the low-momentum scale, describing the center-of-mass momentum of the pair with respect to the rest of the nucleus, on the order of 100--150~MeV/$c$. The differential cross section takes the factorized form
\begin{equation}
     d\sigma = \sigma_{ep} \cdot K \cdot \sum_{a} C_a \cdot \mathcal{P}_{\rm{rel}}^a
    \cdot \mathcal{P}_{\rm{CM}}^a,
\end{equation}
where $\sigma_{ep}$ is the cross section for an electron to scatter from an off-shell proton in the nucleus; $K$ is a kinematic factor that depends on the final state momentum vectors; $a$ represents the possible quantum numbers for an SRC pair; $C_a$ is the contact term, i.e., the abundance of pairs with quantum numbers $a$ in the nucleus; $\mathcal{P}_{\rm{rel}}^a$ is a distribution governing relative motion of the nucleons in a pair of type $a$; and $\mathcal{P}_{\rm{CM}}^a$ is a distribution governing the center-of-mass motion of a pair of type $a$. For the $(e,e'pp)$ reaction, in which the struck nucleon is a proton, and the correlated spectator is also a proton, the sum over $a$ only has one term; $pp$ pairs with spin-0 (Pauli exclusion suppresses spin-1 pairs). For the $(e,e'pn)$ reaction, the sum over $a$ includes $pn$ spin-1 and $pn$ spin-0 pairs. There are multiple approaches to specifying $\mathcal{P}_{\rm{rel}}^a$ and $\mathcal{P}_{\rm{CM}}^a$, depending on assumptions made. Both the center-of-mass and relative distributions depend on the initial momentum (prior to the scattering reaction) of the struck nucleon, $\vec{p}_i$, and the initial momentum of the spectator nucleon, $\vec{p}_2$. We define
\begin{align*}
    \vec{p}_{\rm{CM}}&=\vec{p}_i+\vec{p}_2\\
    \vec{p}_{\rm{rel}}&=\frac{\vec{p}_i-\vec{p}_2}{2}
\end{align*}
as the center-of-mass momentum vector of the pair and the relative momentum vector of the pair, respectively.

In this work we adopt the ``light-cone'' GCF specification of Ref.~\cite{Pybus2020}, since it more naturally handles relativistic effects, which are likely to be important for nucleons moving with momenta approaching 1~GeV$/c$. In light cone coordinates, four-momentum vectors $(p^0,p^1,p^2,p^3)$ are represented in terms of transverse components, $\vec{p}_{\perp} \equiv (p^1,p^2)$, as well as plus-momentum and minus-momentum components $p^{\pm} \equiv p^0 \pm p^3$.
We use $\alpha \equiv p^-/\tilde{m}$, where $\tilde{m}=m_A/A$ and $m_A$ is the mass of the target nucleus, to represent the fractional minus-momentum; for example, we use $\alpha_{\rm{CM}}$ to represent the fractional minus-momentum of the center-of-mass of the pair. In the light-cone GCF specification, the 8-fold differential cross section for the $(e,e'pN)$ reaction can be written as
\begin{equation}
\frac{d^8\sigma}{d\Omega_{k'}d^3\vec{p}_{\rm{CM}}d^3\vec{p}_{\rm{rel}}} = 
\sigma_{ep} \cdot \left[ \frac{1}{{4\pi\alpha_i} \lvert 1-\frac{\vec{p}_i \cdot \vec{k}'}{E_i E'_k} \rvert}  \right] 
    \cdot \sum_{a} C_a \cdot \mathcal{P}_{\rm{rel}}^{\rm{LC},a}
    \cdot \mathcal{P}_{\rm{CM}}^{\rm{LC},a},
    \label{eq:gcf_cs}
\end{equation}
where the outgoing electron momentum, solid angle, and energy are given by $k'$, $\Omega_{k'}$, and $E_{k'}$, respectively, and $\alpha_i$ is the initial minus-momentum fraction of the struck nucleon prior to the collision. Ref.~\cite{Pybus2020} gives a light-cone expression for $\mathcal{P}_{\rm{rel}}^{\rm{LC},a}$ as
\begin{equation}
    \mathcal{P}_{\rm{rel}}^{\rm{LC},a} = \frac{1}{E_2}\cdot \frac{\sqrt{m_N^2 + k^2}}{2-\alpha_{\rm{rel}}}\cdot \frac{\lvert \tilde{\phi}^a(k)\rvert^2}{(2\pi)^3}
\end{equation}
    where $\lvert \tilde{\phi}^a(k)\rvert ^2$ is the relative momentum distribution for nucleons in an SRC pair with quantum numbers $a$, which should be universal across nuclei; $m_{N}$ is the nucleon mass; $E_2=\sqrt{\vec{p}^{\ 2}_2+m^2_N}$ is the energy of the spectator nucleon in the pair; and $k$ is given by 
   \begin{equation}
        k^2 \equiv \frac{m_N^2 + p_{\rm{rel},\perp}^2}{\alpha_{\rm{rel}}(2-\alpha_{\rm{rel}}) } - m_N^2.
    \end{equation}
Following Ref.~\cite{Pybus2020}, we assume that the center-of-mass distribution is well-modelled by a Gaussian distribution in light-cone coordinates,
\begin{equation}
    \mathcal{P}_{\rm{CM}}^{\rm{LC},a} = \frac{\tilde{m} \alpha_{A-2}}{E_{A-2} \left(2\pi\sigma_{\rm{CM}}^{a\ 2} \right)^{3/2} }
        \exp \left\{ -\frac{\tilde{m}^2(2-\alpha_{\rm{CM}})^2 + {\vec{p}_{\rm{CM},\perp}}^{\ 2}}{2{\sigma_{\rm{CM}}^{a\ 2}}} \right\},
\end{equation}
with a single parameter $\sigma_{\rm{CM}}^a$ governing the width of the distribution. Here, $E_{A-2}$ and $\alpha_{A-2}$ represent the energy and minus-momentum fraction, respectively, of the residual $A-2$ system. 

For our simulations, we produced Monte Carlo events weighted according to the cross section specified in Eq.~\ref{eq:gcf_cs}. More details on GCF and its implementation as a Monte Carlo event generator can be found in Ref.~\cite{Pybus2020}.

% GCF inputs
The GCF requires several model inputs, and for consistency with respect to previous calculations~\cite{Korover2021,CLAS:2020mom}, we have matched their choices of values for those inputs. The inputs include the contact $C_a$ for each set of nucleon pair quantum numbers, obtained theoretically and in agreement with experiment \cite{Weiss2018}; the width of the center-of-mass momentum distribution, $\sigma_{\rm{CM}}^a$, which we take to be the same for various SRC analyses with various targets
~\cite{CLAS:2020mom,Korover2021}, with a value of 150$\pm$20 MeV/$c$~\cite{Cohen2018}; the residual excitation energy $E^*$ for the $A-2$ system, which we varied uniformly from 0-30 MeV following \cite{Korover2021,CLAS:2020mom} (which based their estimate on Ref.~\cite{Ulmer1987});
and a model $NN$ interaction used to determine the universal two-nucleon wave functions $\tilde\phi^{\alpha}(k)$, for which we use the AV18 model~\cite{Wiringa1995}. GCF has so far proven quite effective in predicting a variety of nuclear properties \cite{Weiss2018,Pybus2020,CLAS:2020mom,Korover2021,Cruz-Torres2021,Weiss2019,Cohen2018}.

Of course, reaction mechanisms not accounted for in the formalism described above also occur. These effects may include elastic and inelastic final state interactions and recoil events in which a knocked-out neutron interacts with other nucleons before leaving the nucleus \cite{CLAS:2020mom}. The Glauber approximation is used to estimate the probability of FSI under different kinematic conditions and to establish event selection criteria to choose events with a low probability of FSI. We therefore chose to conduct our study in antiparallel kinematics, in which the angle between the missing momentum $\vec{p}_{\rm{miss}}$ and electron momentum change $\vec{q}$ is close to 180$^{\circ}$. In Refs. \cite{Hen:2014nza,Cohen2018,Duer2018,Duer2019,LabHallA:2014wqo,Korover2021}, suppression of non-plane wave reaction mechanisms in antiparallel kinematics was accomplished by a common set of event selection criteria or ``cuts." The contribution from some reaction effects, including SCX, are somewhat more difficult to estimate, underscoring the importance of the $pn/p$ measurement.

To account for the changes in the experimental cross sections due to nuclear transparency and SCX, we used a simplified version of the procedure described in Ref.~\cite{Pybus2020} and in the Supplementary Information of Ref.~\cite{CLAS:2020mom}. Because the transparencies and SCX probabilities given in Ref.~\cite{Colle2016} are fairly rough estimates, we neglect for the time being other kinematic-dependent effects, such as the decrease in the effect of SCX with increasing momentum of the struck proton~\cite{Gibbs1994}.

The precise effect of SCX on the $(e,e'p)$ and $(e,e'pN)$ reactions is difficult to calculate. We estimate the effect of SCX on the measured cross sections, which we denote by $(e,e'p)_{\rm{SCX}}$ and $(e,e'pN)_{\rm{SCX}}$, by forming a mixture of the $(e,e'pp)$ and $(e,e'pn)$ channels predicted by our GCF simulations. Glauber calculations~\cite{Colle2016} performed to support measurements in similar kinematics~\cite{DuerPRL,CLAS:2020mom} found single-charge exchange probabilities for leading and recoiling nucleons being ejected from carbon to be on the order of 4--5\%. Assuming, for simplicity, that the probability of SCX, $P_{\text{SCX}}$, is the same for leading and recoiling nucleons, and for both protons and neutrons, with a value of 5\%,
we find
\begin{align}
    (e,e'pp)_{\rm{SCX}} &= (1-2P_{\text{SCX}}) \cdot (e,e'pp) + \left( 1+ \frac{\sigma_{en}}{\sigma_{ep}} \right) \cdot (e,e'pn)\cdot P_{\rm{SCX}} \label{eq:scx1}\\
    (e,e'pn)_{\rm{SCX}} &= (1-2P_{\text{SCX}}) \cdot (e,e'pn) + \left( 1+ \frac{\sigma_{en}}{\sigma_{ep}} \right) \cdot (e,e'pp) \cdot P_{\rm{SCX}} \label{eq:scx2}\\
    (e,e'p)_{\rm{SCX}} &\approx (e,e'p) \label{eq:scx3}
\end{align}
where $\sigma_{ep}/\sigma_{en}$ is the ratio of electron-proton to electron-neutron cross sections. In the kinematics of this experiment,  $\sigma_{ep}/\sigma_{en} \approx 2.5$, according to the nucleon form factor parameterization of Ref.~\cite{Kelly:2004hm}.
We neglect the effect of SCX on the $(e,e'p)$ cross section, since it is relatively small compared to that on $(e,e'pn)$ and $(e,e'pp)$.
A derivation of the above equations can be found in the Appendix. Given the aforementioned numerical values, $(1-2P_{\text{SCX}})$ evaluates to 0.9, and $(1+ \frac{\sigma_{en}}{\sigma_{ep}})P_{\text{SCX}}$ is 0.07.

We modeled the attenuation of outgoing nucleons by the nuclear medium as a constant transparency factor, with values taken from Ref.~\cite{Duer_thesis,CLAS:2020mom} based on the Glauber calculations described in Ref.~\cite{Colle2016}. In this context, nuclear transparency is the probability that an outgoing proton or neutron emerges from the nucleus rather than failing to emerge due to scattering from the surrounding nuclear material. We used a transparency value of $T^{N}$=0.53 for the $(e,e'p)$ yields to indicate the probability that the leading proton escapes the nucleus, and we used a value of $T^{NN}$=0.44 for the $(e,e'pN)$ yields to indicate that both the leading proton and recoil nucleon escape. These transparency factors were applied to the eight cross sections given in Eqs.~\ref{eq:scx1}-\ref{eq:scx3}.

\section{Results}
\label{sec:results}

\subsection{Fast Monte Carlo Event Selection}
\label{ssec:fastmc}

\begin{figure}[h]%
\centering
\includegraphics[width=0.7\textwidth]{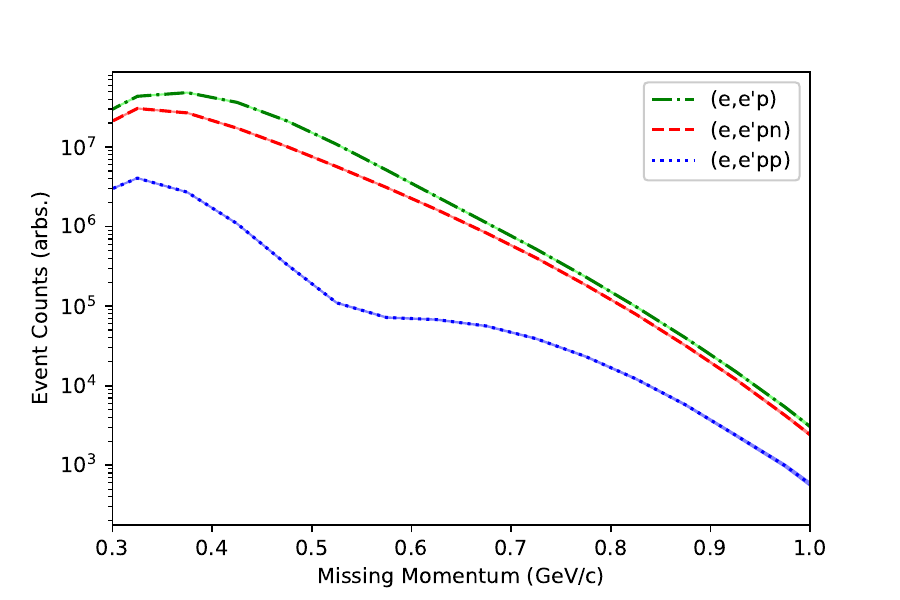}
\caption{Distribution of the missing momentum for $(e,e'p)$, $(e,e'pn)$, and $(e,e'pp)$ for an arbitrary number of physics events.}
\label{fig:fastmc}
\end{figure}

To understand the distribution of events without any influence from experimental or detector effects, we first performed a ``Fast Monte Carlo'' analysis, integrating carbon events from the Monte Carlo event generator at a beam energy of 6 GeV over bounds defined by the approximate acceptance of the detector with an approximate set of event selection criteria. We stratified these events over the different sets of selection criteria, labeling them as $(e,e'p)$, $(e,e'pp)$, and $(e,e'pn)$, depending on which criteria are passed. The definitions of the criteria are explained below. The missing momentum distribution for each channel, based on the Fast Monte Carlo analysis, is shown in Fig.~\ref{fig:fastmc}.

For this study, we require that leading protons pass through the Forward Detector rather than the Central Detector by requiring their polar angle to be $\theta_p<$ 40$^{\circ}$. This requirement is partly due to the limited momentum resolution of the Central Detector, which limits the experimental resolution of $p_{\rm{miss}}$ for Central Detector protons. Additionally, our simulation indicates that at least as many leading protons will pass through the Forward Detector as through the Central Detector, making forward protons advantageous for high event counts.

\begin{figure}[h]%
\centering
\includegraphics[width=0.7\textwidth]{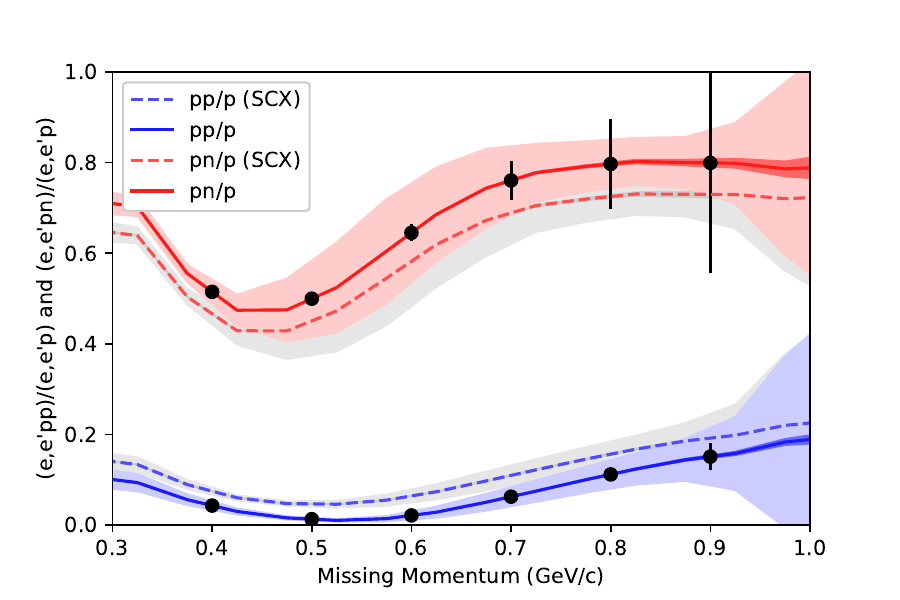}
\caption{
\label{fig:pNp}
Estimates of the $(e,e'pn)/(e,e'p)$ (solid red) and $(e,e'pp)/(e,e'p)$ ratios (solid blue) for simulated SRC events. The dark error bands indicate simulation statistical error, the light error bands indicate systematic error due to uncertainty in nuclear properties, and the black error bars indicate the expected statistical error from the Nuclear Targets Experiment. The dotted red and blue lines indicate $(e,e'pn)_{\rm{SCX}}/(e,e'p)$ and $(e,e'pp)_{\rm{SCX}}/(e,e'p)$, respectively, and the associated gray bands indicate their systematic error.}
\end{figure}

\begin{figure}[h]%
\centering
\includegraphics[width=0.7\textwidth]{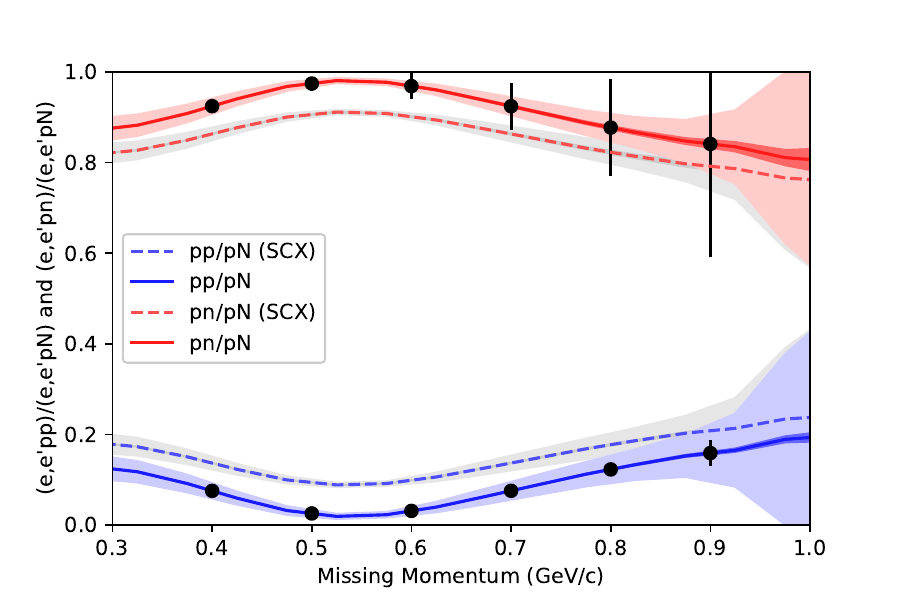}
\caption{
\label{fig:pNp-wrec}
Estimates of the $(e,e'pn)/(e,e'pN)$ (solid red) and $(e,e'pp)/(e,e'pN)$ ratios (solid blue) for simulated SRC events. The dark error bands indicate statistical error, the light error bands indicate systematic error due to uncertainty in nuclear properties, and the black error bars indicate the expected statistical error from the Nuclear Targets Experiment. The dotted red and blue lines indicate $(e,e'pn)_{\rm{SCX}}/(e,e'pN)_{\rm{SCX}}$ and $(e,e'pp)_{\rm{SCX}}/(e,e'pN)_{\rm{SCX}}$, respectively, and the associated gray bands indicate their systematic error.}
\end{figure}

Our $(e,e'p)$ sample consists of leading protons that also meet several additional criteria, constraining the interaction to so-called antiparallel kinematics, in which $\vec{p}_{\rm{miss}}$ is roughly antiparallel to the momentum transfer $\vec{q}$. This is enforced by selecting events with $x_B>1.1$, where $x_B$ is the Bjorken scaling parameter defined by $x_{B}\equiv Q^{2}/2m_{N}\omega$, $\omega$ is the energy transferred by the electron, and $Q^{2}$ is the squared four-momentum transfer given by $Q^2\equiv q^2-\omega^2$. Under the Plane Wave Impulse Approximation (PWIA), we assume that the electron scatters from a single nucleon that leaves the nucleus with momentum $\vec{p}_{1}$ without interaction with other nucleons, so we can approximate the initial proton momentum as $\vec{p}_{i} \approx \vec{p}_{\rm{miss}}$. We accept only leading protons with $p_{\rm{miss}}>$ 0.3~GeV/$c$ to restrict the yields to events with high-momentum nucleons. We also require the angle between $\vec{p}_{1}$ and $\vec{q}$ to be $\theta_{pq}<25^{\circ}$, and we restrict the ratio of their magnitudes to be $0.62 < p_1/q < 1.1$, both so that the struck proton carries a large fraction of the momentum transfer. Finally, we select events with missing mass $M_{\rm{miss}}=\sqrt{(q^{\mu}-p^{\mu}_{1}+p_d^\mu)^{2}}<$ 1.1 GeV/$c^{2}$, where $p_d^\mu=(2m_N,\vec{0})$ in order to suppress contributions from non-quasielastic events, such as those involving pion production and nucleon resonances. These event selection criteria, which are presented succinctly in Table \ref{tab:rgm-cuts}, ensure that we are selecting events with a sufficiently high $Q^2$, that is, at least 1.5~GeV/$c^2$.

\begin{table}[htb!]
\centering
\caption{\label{tab:rgm-cuts} Event Selection Criteria for SRC Events}
\begin{tabular}{|c|c|c|c|}
\hline
$(e,e'p)$ Selection Criteria \\
\hline
$\theta_p <$ 40$^{\circ}$\\
$x_{B} >$ 1.1\\
$p_{\rm{miss}} >$ 0.3 GeV/$c$\\
$\theta_{pq} <$ 25$^{\circ}$\\
0.62 $< p_1/q <$ 1.1\\
$M_{\rm{miss}}< $ 1.1 GeV/$c^2$\\
\hline
\end{tabular}
\end{table} 

Finally, the $(e,e'pN)$ sample contains the subset of $(e,e'p)$ events with a recoil nucleon, either a proton or a neutron, with a momentum larger than 0.3~GeV/$c$ and a polar angle $\theta_{\text{rec}}$ in the range $40^{\circ}<\theta_{\text{rec}}<140^{\circ}$, which is the approximate coverage of the Central Detector, through which our simulations have indicated that most recoil nucleons will pass.

Our final observables, the resulting $pn/p$ and $pp/p$ ratios, are shown in Fig.~\ref{fig:pNp} with their statistical and systematic errors. The statistical error bands are not meant to reflect expectations for the Nuclear Targets Experiment but to show convergence of the GCF prediction. We simulated a sufficiently high number of events in order to make the simulation statistical error negligible compared to the systematic error and in order to obtain a more precise value of the high-$p_{\rm{miss}}$ limit of the ratios. The systematic error was obtained by taking the standard deviation of the $pp/p$ and $pn/p$ ratios for several hundred simulations with randomized nuclear properties, namely $C_a$, $E^*$, and $\sigma^a_{CM}$. We also show the $(e,e'pn)_{\rm{SCX}}/(e,e'p)$ and $(e,e'pp)_{\rm{SCX}}/(e,e'p)$ ratios as an estimate of how the $pn/p$ and $pp/p$ may deviate from the predictions of GCF given the effects of single charge exchange. $(e,e'pn)_{\rm{SCX}}/(e,e'p)$ and $(e,e'pp)_{\rm{SCX}}/(e,e'p)$ are shown with their respective systematic error, calculated in the same way as for $pn/p$ and $pp/p$.

\begin{figure}[h]%
\centering
\includegraphics[width=0.7\textwidth]{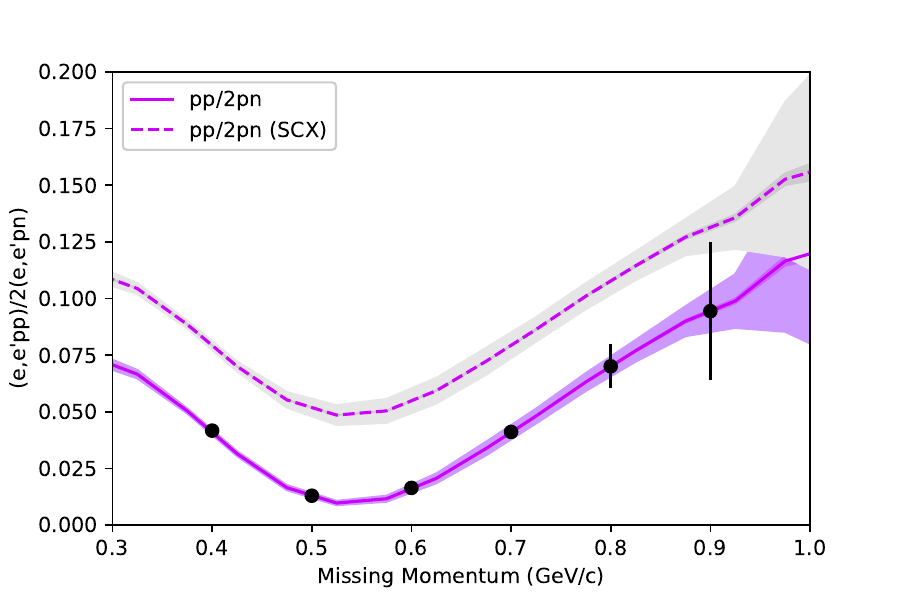}
\caption{
\label{fig:fast_pppn}
Estimates of the $(e,e'pp)/2(e,e'pn)$ ratio (solid purple) for simulated SRC events. The dark error bands indicate statistical error, the light error bands indicate systematic error due to uncertainty in nuclear properties, and the black error bars indicate the expected statistical error from the Nuclear Targets Experiment. The dotted purple line indicates $(e,e'pn)_{\rm{SCX}}/2(e,e'pp)_{\rm{SCX}}$, and the associated gray band indicate its systematic error.}
\end{figure}

Additionally, as an alternative observable, we show in Fig.~\ref{fig:pNp-wrec} the cross section ratios $pp/pN$ and $pn/pN$, that is, $(e,e'pp)/(e,e'pN)$ and $(e,e'pn)/(e,e'pN)$, respectively.
We also present the expected ratio $pp/2pn$ = $(e,e'pp)/2(e,e'pn)$ in Fig.~\ref{fig:fast_pppn}. Both figures also include an estimate of how the observables may change due to SCX.

The black statistical error bars shown in Figs.~\ref{fig:pNp}, \ref{fig:pNp-wrec}, and \ref{fig:fast_pppn} are those we expect for the $^{12}$C data under a 6 GeV beam from the CLAS12 Nuclear Targets Experiment. To obtain the statistical error estimates, the analysis described above, with additional event selection as described in Section \ref{ssec:geant}, was performed on a sample of preliminary $^{4}$He data from the Nuclear Targets Experiment. The resulting $^{4}$He yields were scaled to obtain expected $^{12}$C yields using the total experiment integrated luminosity for $^{4}$He, the relative integrated luminosity between $^{4}$He and $^{12}$C, the nuclear transparency factors for the two nuclei, and their values of $a_2$, a scaling factor which quantifies the abundance of SRC pairs in a given nucleus \cite{Hen2012}.

\subsection{Geant4 Event Selection}
\label{ssec:geant}

\begin{figure}[h]%
\centering
\includegraphics[width=0.7\textwidth]{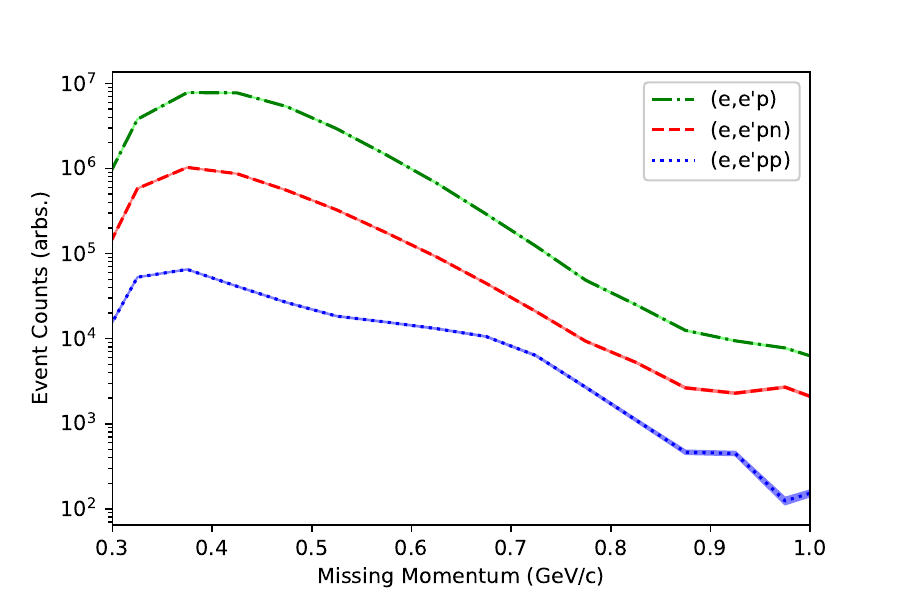}
\caption{
\label{fig:geant}
Distribution of the missing momentum for $(e,e'p)$, $(e,e'pn)$, and $(e,e'pp)$ for an arbitrary number of physics events after CLAS12 event reconstruction, with error bands for statistical error.}
\end{figure}

We have also estimated the carbon $(e,e'p)$, $(e,e'pp)$, and $(e,e'pn)$ missing momentum distributions at a beam energy of 6 GeV for Monte Carlo simulations run through CLAS12's Geant4 program and reconstructed with CLAS12 software. We define these yields based on reconstructed quantities. They include the effects of detector efficiency and resolution and are intended to approximate the results of this analysis on experimental data in the absence of corrections due to the momentum, momentum resolution, and detection inefficiency. The results are shown in Fig.~\ref{fig:geant}.

The event selection for the Fast Monte Carlo yields all apply to analysis of the simulated events reconstructed after Geant4, but the events passing through the simulated detector require that some additional criteria be applied for particle identification.

For electron identification, we require between 1 and 100 photoelectrons to be detected in the High Threshold Cherenkov Counter, and we require electron detection to occur in the fiducial region of the electromagnetic calorimeter. Additionally, we require that the energy deposition match the momentum of the associated electron track by restricting the sampling fraction to be between 0.18 and 0.28. We also require the electron to have momentum in the range $0.5-6$ GeV/$c$. An electron detection efficiency exceeding 99\% is expected~\cite{nima-clas12}.

To identify the leading proton with momentum $p_1$, we require a hit in the Forward Time-of-Flight detector, with leading proton polar angle in the range $\theta_p < 40^{\circ}$. Based on recent upgrades using Machine Learning in particle tracking, we expect at least 98\% proton detection efficiency in the Forward Detector, depending on luminosity~\cite{gavalian1,gavalian2,gavalian3}. We also restrict the difference in the $z$ component of the vertex position between the leading proton and electron to be between -3 and 3 cm. We allow only protons with well-reconstructed momentum by requiring $-0.05 < \Delta \beta < 0.05$, where $\Delta \beta$ is the difference between the reconstructed velocity and $p_{\rm{1}}/ \sqrt{p^{2}_{\rm{1}}+m^{2}_{p}}$, where $m_{p}$ is the proton mass.

For recoil nucleons, we select protons with tracks in the CVT and neutrons with hits in the CND. For both, the momentum must be at least 300 MeV/$c$, and the polar angle must fall between 40$^{\circ}$ and 140$^{\circ}$. For recoil protons, we expect the detection efficiency to drop from a maximum of 75-80\% for protons with a momentum of 1 GeV/$c$ to a negligible efficiency at about 0.3 GeV/$c$~\cite{nima-clas12}. As previously mentioned, we expect a neutron detection efficiency of up to 10\%~\cite{nima-clas12}.

These criteria are meant to be minimal and they lead to a negligible reduction in yield. In the analysis of data, in which various backgrounds may be of concern, more stringent criteria may become necessary.

\section{Discussion}
\label{discussion}

The theoretical predictions presented here differ slightly from previous predictions of the tensor-to-scalar transition, primarily due to detector acceptance. Our results from Section \ref{ssec:geant} are not acceptance- or efficiency-corrected, i.e., they reflect predictions for the raw yield measured in the experiment under CLAS12 reconstruction.
Ref.~\cite{CLAS:2020mom} similarly  presented non-corrected GCF predictions for an analysis of data from CLAS, which has a different acceptance, particularly for recoil nucleons. Ref.~\cite{Korover2021} performed an acceptance correction, though with large systematic uncertainties. Older works, such as Ref.~\cite{LabHallA:2014wqo} predate the development of GCF and instead used a simple model of the abundance and center-of-mass distribution of SRC pairs in the nucleus.

We have neglected several factors that will be important in making the measurement. First, the proton tracking efficiency in the CLAS12 Central Vertex Tracker and the neutron detection efficiency in the Central Neutron Detector are the largest sources of systematic uncertainty, especially for the $pn/p$ measurement. Second, the Forward Detector proton momentum resolution is expected to be on the order of 1\%, while the Central Detector proton momentum resolution is expected to be on the order of 5\%. The Central Detector neutron momentum resolution has not yet been precisely measured. Finally, the geometry and imperfect proton efficiency of the Central Detector allow improperly tracked protons to be mis-reconstructed as neutrons, leading to confusion between the $(e,e'pp)$ and $(e,e'pn)$ samples.

A thorough understanding of the proton and neutron efficiencies and momentum resolutions will be crucial to the analysis, as will a method to reject charged particle contamination in the neutron sample. These studies are already underway, with promising results. With corrections made for these effects, the estimates on our statistical uncertainty indicate that we are likely to be able to measure the change in the $pn/p$ ratio. The results reported in this study offer a promising outlook on the possibility of measuring the transition from a tensor-dominated $NN$ interaction at 300--600~MeV/$c$ to a scalar-dominated interaction at high missing momentum using the CLAS12 detector.

\backmatter

\bmhead{Acknowledgments}

This work was supported by the US Department of Energy Office of Science, Office of Nuclear Physics, under contract no. DE-SC0016583; a Columbian Fellowship from The George Washington University; and a JSA/JLab Graduate Fellowship. The authors thank Jackson Pybus for the use of his GCF event generator code.

\bmhead{Data availability statement}

This manuscript has no associated data or the data will not be deposited. The physics events used in this analysis are simulated using Generalized Contact Formalism cross sections as described in the text. Further information about the generator code is available upon reasonable request.

\section*{Declarations}

\bmhead{Conflict of interest}

The authors declare that they have no known competing financial interests or personal relationships that could have appeared to influence the work reported in this paper.

 \begin{appendices}

\section{Simple Estimate of Single Charge Exchange}\label{app:scx}

Single charge exchange can result in an SRC pair being detected in an isospin configuration other than the one in which it occurred when the leading nucleon is struck by the electron. For example, a $pn$ pair can be detected as a $pp$ pair, an $nn$ pair as a $pn$ pair, etc. Estimates based on the Glauber approximation have found approximately a 4-5\% probability of single charge exchange in CLAS kinematics, depending on the pair's isospin configuration~\cite{Colle2016,DuerPRL,CLAS:2020mom}. For simplicity, we use 5\% for all pairs, and we neglect the possibility of both nucleons in the pair undergoing SCX. Here we derive our simple estimate found in Eqs.~\ref{eq:scx1},~\ref{eq:scx2}, and~\ref{eq:scx3}. 

The number of $pp$-pairs observed in an experiment, $N_{pp}^\text{exp.}$, can be written
\begin{equation}
            \label{eq:pp_SCX}   
    N_{pp}^{\rm{exp.}} = N_{\rm{SRC}}\cdot\left[R_{pp}\cdot 2\sigma_{ep}\cdot(1-2P_\text{SCX}) +(1-2R_{pp})\cdot(\sigma_{ep}+\sigma_{en})P_{\rm{SCX}}\right],
\end{equation}
    where $R_{pp}$ is the fraction of SRC pairs that are $pp$ pairs, $P_{\rm{SCX}}$ is the probability for a
    nucleon (either leading or recoil) to undergo single charge exchange, $\sigma_{ep}$ and $\sigma_{en}$ are respectively the off-shell electron-proton and electron-neutron cross sections, and $N_{\rm{SRC}}$ is an overall normalization factor.
    This expression assumes that the number of $pp$ pairs and the number of $nn$ pairs are equal (in other words, $2R_{pp}+R_{pn}=1$) and neglects the simultaneous charge exchange of both leading and recoil nucleons. The first term of Eq.~\ref{eq:pp_SCX} represents scattering from a true $pp$ pair, while the second term represents scattering from either the $p$ or the $n$ of an $np$ pair, which then undergoes single charge exchange so as to appear like a $pp$ pair in the experiment. If there were no single charge exchange, i.e. $P_{\rm{SCX}}=0$, then the experiment would have instead observed
    \begin{equation}
            \label{eq:pp_no}
    N_{pp}^\text{no SCX} = N_{\rm{SRC}}\cdot R_{pp}\cdot 2\sigma_{ep}.
\end{equation}

We can write a similar expression for the number of $np$-pairs observed through the $(e,e'pn)$ reaction, i.e., with a leading proton, in an experiment,
    \begin{equation}
                \label{eq:pn_SCX}   
        N_{pn}^{\rm{exp.}} = N_{\rm{SRC}}\cdot\left[ (1 -2R_{pp})\cdot\sigma_{ep}\cdot(1-2P_\text{SCX}) + 2 R_{pp}\cdot(\sigma_{ep}+\sigma_{en})\cdot P_\text{SCX} \right].
    \end{equation}

\noindent In the limit of no single charge exchange, this reduces to
        \begin{equation}
        \label{eq:pn_no}
        N_{pn}^\text{no SCX} = N_{\rm{SRC}}\cdot (1 -2R_{pp})\cdot\sigma_{ep}.
    \end{equation}

By substituting Eqs.~\ref{eq:pp_no} and \ref{eq:pn_no} into Eqs.~\ref{eq:pp_SCX} and \ref{eq:pn_SCX}, we find
\begin{align}
    N_{pp}^{\rm{exp.}} &= N_{pp}^\text{no SCX} (1-2P_\text{SCX}) + N_{pn}^\text{no SCX} \left( 1+ \frac{\sigma_{en}}{\sigma_{ep}} \right) P_\text{SCX} \\
    N_{pn}^{\rm{exp.}} &= N_{pn}^\text{no SCX} (1-2P_\text{SCX}) + N_{pp}^\text{no SCX} \left( 1+ \frac{\sigma_{en}}{\sigma_{ep}} \right) P_\text{SCX}.
\end{align}
These expressions are used for the numerical estimates given in Eqs.~\ref{eq:scx1} and \ref{eq:scx2}.

Adding together Eqs.~\ref{eq:pp_SCX} and~\ref{eq:pn_SCX}, we obtain an expression for the expected number of observed $(e,e'p)$ SRC pairs,

    \begin{equation}
        N_{p}^{\rm{exp.}} = N_{\rm{SRC}} \cdot [\sigma_{ep}(1-2P_{\rm{SCX}}) + (\sigma_{en}+\sigma_{ep})P_{\rm{SCX}}].
    \end{equation}
\\
\noindent We compare $N_{p}^{\rm{exp.}}$ to the number of $(e,e'p)$ pairs expected in the absence of SCX, namely, $N_{p}^{\text{no SCX}} = N_{\rm{SRC}}\sigma_{ep}$, by examining the fractional difference between the two,

    \begin{equation}
        \label{eq:p_Delta}
        \frac{\Delta_{p}}{N_{p}^{\text{no SCX}}} = \frac{N_{p}^{\text{exp.}} - N_{p}^{\text{no SCX}}}{N_{p}^{\text{no SCX}}} = P_{\rm{SCX}}\left( \frac{\sigma_{en}}{\sigma_{ep}} - 1 \right).
    \end{equation}
\\
\noindent Numerically, Eq.~\ref{eq:p_Delta} evaluates to -0.03. We therefore make the approximation that the change in the $(e,e'p)$ cross section due to SCX is negligible.
    
\end{appendices}

\section*{\protect Bibliography}
\renewcommand\refname{}

\bibliography{sn-bibliography}

\end{document}